\documentclass[reprint,amsmath,amssymb,aps,prb]{revtex4-1}

\usepackage{graphicx,dcolumn,bm,braket,ulem}
\usepackage[usenames, dvipsnames]{color}
\usepackage[pagewise]{lineno}
\begin{document}

\title{Tunable Acoustic Valley-Hall Edge States in Reconfigurable Phononic Elastic Waveguides}

\author{Ting-Wei Liu}
\author{Fabio Semperlotti}
 \email{fsemperl@purdue.edu}
\affiliation{Ray W. Herrick Laboratories, School of Mechanical Engineering, Purdue University, West Lafayette, Indiana 47907, USA.}

\begin{abstract}
This study investigates the occurrence of acoustic topological edge states in a 2D phononic elastic waveguide due to a phenomenon that is the acoustic analogue of the quantum valley Hall effect.
We show that a topological transition takes place between two lattices having broken space inversion symmetry due to the application of a tunable strain field. This condition leads to the formation of gapless edge states at the domain walls, as further illustrated by the analysis of the bulk-edge correspondence and of the associated topological invariants. Although time reversal symmetry is still intact in these systems, the edge states are topologically protected when inter-valley mixing is either weak or negligible. 
Interestingly, topological edge states can also be triggered at the boundary of a single domain if boundary conditions are properly selected. We also show that the static modulation of the strain field allows tuning the response of the material between the different supported edge states.
\end{abstract}
\maketitle
\section{Introduction}
Topological acoustics has rapidly emerged as a new and fascinating branch of physical acoustics. Following the groundbreaking research in solid state physics, which showed the existence of topological states of matter \cite{ReviewKaneTI, ReviewNiu}, researchers have been able to formulate the acoustic analogue of selected mechanisms \cite{TopologicalAcoustics-Flow, MechanicalTI, TopoPhon-Gyro, TopoSound-Flow, AcousticTIPlate, TopoGyroExp, ValleySonicBulk, TunableTopoPnc-Flow, AcousticTIAir, ValleySonicEdge, ValleySonicEdge}. 
During the last decade, several studies have investigated topological materials based on broken space-inversion symmetry (SIS). 
In electronic materials, SIS was broken using different methodologies including graphene-like lattices with staggered sublattice potentials \cite{ValleyContrasting, DomainWall, ControlEdgeStates, SiliceneBrokenSIS,SiliceneDW}, strained graphene \cite{StrainedGraphene}, and multi-layered graphene under electric fields \cite{EdgeStatesMultilayerGraphene,MultilayerGraphene,BilayerGrapheneDW, ValleyChernBilayerGraphene,BilayerGraphene,Highway,GateControl}. The effect of the asymmetry is that of opening a gap at the original Dirac cone associated with the hexagonal lattice in which edge states are now supported. These edge states cannot be explained by the quantum Hall effect (QHE) mechanism. In fact,
as TRS is intact
the lattice still possesses a trivial topology within the context of QHE \cite{ReviewKaneTI, Raghu, PhotonicGraphene}. However, due to the large separation in $\mathbf{k}$-space of the two valleys, valley-dependent topological invariants can be defined and used to classify the topological states of the different lattices. This approach, usually referred to as quantum valley Hall effect (QVHE), was recently investigated for application to fluidic acoustic waveguides \cite{ValleySonicBulk, ValleySonicEdge}, as well as elastic plates with local resonators \cite{Ruzzene1, Ruzzene2}.

\begin{figure}[ht]
\includegraphics[scale=0.83]{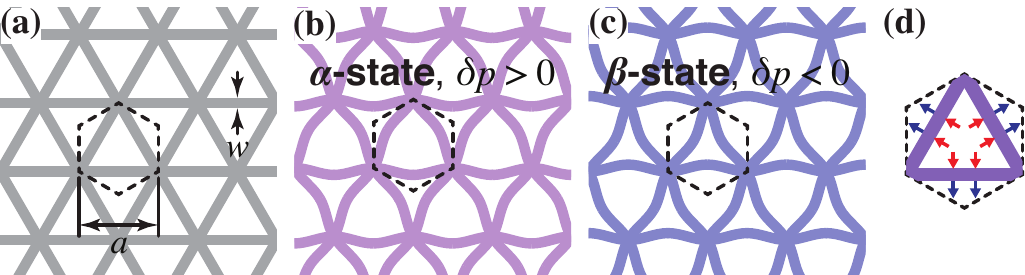}
\caption{\label{fig:lattice} Schematic illustration of the lattice geometry. (a) The reference crystal (unstrained) consisting of an aluminum truss-like hexagonal lattice having SIS. The dashed black line indicates the primitive Wigner-Seitz cell. (b, c) The deformed $\alpha$- and $\beta$-lattices obtained upon application of internal pressure. In both configurations SIS is broken. (d) The red and blue arrows show the equivalent forces produced by $\delta p$. To improve the visualization, the deformation shown in (b), (c) is magnified by a factor of 5 with respect to the actual deformation produced by $\delta p = 20 \mbox{ MPa}.$}
\end{figure}

In the present study, we consider the dispersion and propagation behavior of a topological elastic phononic waveguide assembled from a truss-like unit cell (Fig.~\ref{fig:lattice}(a)). Compared to the previous studies \cite{Ruzzene1, Ruzzene2}, we take a fully continuum modeling approach which provides a general methodology of analysis and allows mapping the topological behavior all the way back to the massless (or massive when SIS is broken) Dirac equation. Also, an in-depth study of the occurrence of edge states, either at domain walls or at the lattice boundaries, is presented. The analysis addresses both the topological significance of the different edge states as well as their lossless behavior. Further, despite the weak topological nature of the QVHE, we show that selective valley injection can be obtained by a dual-excitation approach able to target specific eigenstates. In addition, we show tunable and reconfigurable capabilities of the topological medium. Starting from an initial hexagonal $D_{6h}$ lattice symmetry, the symmetry level can be tuned using an external static actuation which allows real-time tuning of the edge states (i.e. of the corresponding topological bandgap) as well as a complete reconfiguration of the lattice from a regular phononic material to a tunable topological medium, and viceversa. Note that, in this study, we are mostly concerned with the analysis of the physical behavior of the medium and of its tuning capability and we do not concentrate on the details of implementation of the tuning mechanism. Nevertheless, we note that the local perturbation could be practically implemented by using, as an example, mechanical or thermal loads. In the following, we assume that an isostatic pressure could be applied on the side walls to produce the deformation necessary for the topological phase transition.

The individual trusses forming the lattice are assumed made out of aluminum, and having width $w=0.1a$, and thickness $t=0.05a$ where $a$ is the lattice constant. The fundamental lattice structure clearly exhibits SIS with inversion axes located at the nodal points. The lattice is then deformed by applying an internal pressure $\delta p$ that lowers the symmetry to $D_{3h}$. The equivalent force produced by the $\delta p$ is locally normal to the individual truss element as shown in Fig.~\ref{fig:lattice}(d).
The application of either a positive or a negative pressure $\delta p$ results in two different deformed states of the lattice that for simplicity we label as $\alpha$- and $\beta$-states (Fig.~\ref{fig:lattice}(b),(c)), respectively. In these states, the SIS is broken and the states $\alpha$ and $\beta$ are inverted images of each other, that is they turn into each other as $\delta p$ switches sign.

We will use a combination of theoretical and numerical tools to show in the following that 1) the evolution from $\alpha$- to $\beta$-state (and viceversa) is accompanied by a topological phase transition, and that 2) when a phononic system is obtained by assembling the two phases, an edge state is supported along the domain wall (DW, i.e. the interface between the $\alpha$- and $\beta$-state). We note that these edge states can be created and modulated at the boundary of a single lattice by selecting proper boundary conditions and by tuning the internal pressure.

\section{Topological band structure analysis}
\subsection{Phononic band structure}
For a unit cell in prestrained conditions, the numerical analysis can be divided in two parts: 1) a nonlinear static analysis to calculate the state of displacement $\mathbf{u}_0$, stress $\mathbf{s}_0$, and strain $\bm{\epsilon}_0$
induced by the applied pressure load, and 2) a linear wave propagation analysis around the pre-stressed equilibrium state.

In our analysis, the pre-stressed state was calculated using finite element analysis and used as input for the linear wave analysis. The wave propagation in the pre-stressed medium can be studied as a linear small oscillations problem
$\mathbf{u}'e^{i\omega t}$ around the new static equilibrium
$\mathbf{u}_0$. The total displacement can then be expressed as $\mathbf{u(t)}=\mathbf{u}_0+\mathbf{u}'e^{i\omega t}$. Under these conditions the linearized wave equation is given by
\begin{equation} \label{elasticity}
\rho \frac{\partial^2 \mathbf{u}}{\partial t^2}  = -\nabla \cdot \mathbf{P},
\end{equation}
which is readily rewritten as:
\begin{equation} \label{elasticity2}
-\rho \omega^2 \bf u' = -\nabla \cdot \mathbf{P},
\end{equation}
where $\rho$ is the mass density, $\mathbf{P}=(\nabla \mathbf{u} + \mathbf{I})\mathbf{s}$ is the first Piola-Kirchhoff stress tensor,  $\mathbf{s}$ is the second Piola-Kirchhoff stress tensor with $\mathbf{s}=\mathbf{s}_0+\mathbf{C}:(\bm{\epsilon}-\bm{\epsilon}_0)$, where $\mathbf{C}$ is a $4^\textrm{th}$ order elasticity tensor, ${\mathbf{s}_0}$ and $\bm{\epsilon}_0$ are the initial stress and strain, respectively, and the differential operator $\nabla$ is taken with respect to the material frame.
Solving the $\mathbf{k}$-dependent Bloch eigenvalue problem yields the phononic band structure and the eigenstates. The above described model was assembled and solved using the commercial finite element software COMSOL Multiphysics.

The frequency dispersion is normalized by the bulk shear wave speed in aluminum $c$ over the lattice constant $a$. Note that the current system is effectively a flat waveguide that admits Lamb guided modes (symmetric $S$ and antisymmetric $A$), as well as shear horizontal $SH$ modes. In this study, we concentrate only on $A$ modes (flexural modes) that are those more naturally excited in plate-like structures under external excitation. Other modes are filtered out based on particle motion polarization of the mode shapes.
The value of the pressure perturbation $\delta p = 20 \mbox{ MPa}$ was chosen in order to achieve large deflections without inducing plastic deformations. Note that, because the lattice was designed based on slender members, it can easily accumulate large deflections while maintaining small strain levels.

\begin{figure}[ht]
\includegraphics[scale=0.78]{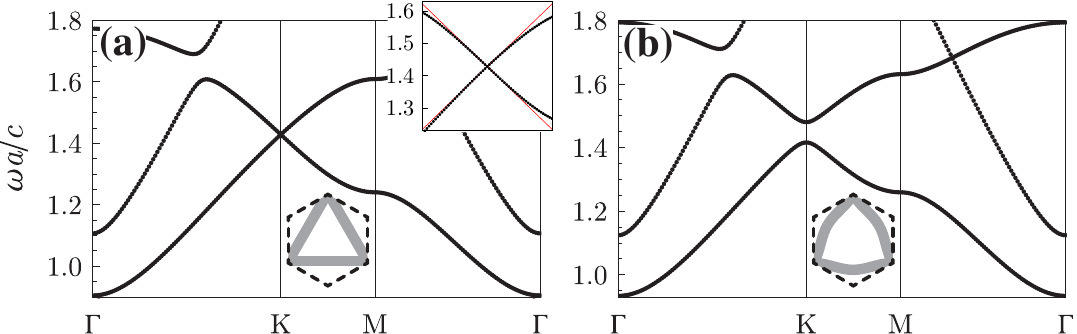}
\caption{\label{fig:bndstr} Phononic band structure of (a) the undeformed lattice (shown in Fig.~\ref{fig:lattice} (a)), and of (b) the $\alpha$- (or $\beta$-) lattice (shown in Fig.~\ref{fig:lattice} (b) and (c)). An incomplete bandgap opens up at the original Dirac point as SIS is broken. The inset in (a) shows the comparison between the numerical dispersion data (black dots) and the dispersion predicted by the $\mathbf{k}\cdot\mathbf{p}$ method (red lines) near the Dirac point.}
\end{figure}

Fig.~\ref{fig:bndstr} (a) and (b) show the band structure of the flexural modes
of the lattice with $\delta p = 0$ (intact SIS) and $\delta p = 20$ MPa (broken SIS), respectively. Note that the two lattices in states $\alpha$ and $\beta$ have identical band structure (Fig.~\ref{fig:bndstr} (b)) since they are simply spatially inverted versions of each other and with intact TRS. 

The analysis of the dispersion properties in Fig.~\ref{fig:bndstr} (a) reveals the existence of a degeneracy at the $\mathbf{K}$ point from which locally linear dispersion curves emanate. This dispersion structure, known as the Dirac cone (DC) \cite{E.Properties.Graphene}, is protected by the lattice configuration. However, when SIS is broken (Fig.~\ref{fig:bndstr} (b)), the degeneracy is lifted and the initially degenerate modes can separate and give rise to a local bandgap.
It is the adiabatic evolution of this specific dispersion structure under small perturbations that enables the generation of topological edge modes. We will show that the dispersion near the valleys in either the reference or the perturbed configuration can be mapped into a massless or a massive Dirac equation, respectively. Before proceeding, we should also note that the current configuration produces an incomplete bandgap (Fig.~\ref{fig:bndstr}b). In principle, a complete bandgap could be obtained by optimizing the lattice geometric parameters \cite{AcousticTIPlate}, however we will show that the valley-dependent design is fairly robust and does not strictly require a full bandgap.

\begin{figure}[ht]
\includegraphics[scale=0.77]{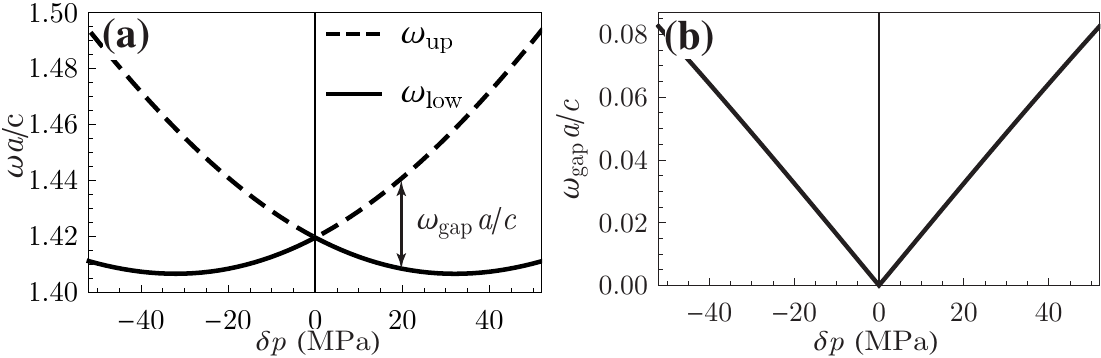}
\caption{\label{fig:fvspr} Effect of the applied perturbation $\delta p$ on (a) the evolution of the upper and lower bounds of the bandgap, and (b) the gap width.}
\end{figure}

As established above, the process of lifting the degeneracy and opening the bandgap is connected to the applied pressure perturbation within the cell. Fig.~\ref{fig:fvspr} (a) and (b) show the evolution of the gap bounds as well as of the width of the gap as a function of the applied pressure $\delta p$. An average upward shift of the frequency is observable in the gap bounds as $\lvert \delta p \rvert$ increases. This behavior is directly related to the stress stiffening effect produced by the geometric nonlinear deformation of the cell. The analysis of the gap width indicates that the gap vanishes and reopens as $\delta p$ crosses zero, therefore suggesting a topological phase transition in the $\delta p$-space. 
\subsection{Berry curvature and valley Chern number}
The evolution of the bandgap as a function of $\delta p$ suggests that, a lattice obtained by connecting $\alpha$ and $\beta$ domains should experience a topological phase transition associated with a vanishing gap occurring exactly at the DW. The topological nature of this transition can be characterized using a topological invariant, that is the Chern number $C_n$ (module $2\pi$). The parameter $C_n$ is obtained by integrating the Berry curvature $\Omega _n(\mathbf{k}) = \nabla _\mathbf{k} \times \braket{\mathbf{u}_n(\mathbf{k})|i\nabla _\mathbf{k}|\mathbf{u}_n(\mathbf{k})}\cdot \hat{\mathbf{z}}$ of the $n^\textrm{th}$ mode throughout the first Brillouin zone. For our system, $C_n$ is expected to be zero due to an odd distribution of the Berry curvature in $\mathbf{k}$-space, which should be expected given TRS is preserved \cite{ReviewKaneTI, Raghu, PhotonicGraphene}. It follows that these lattices are classified as trivially gapped materials in the context of QHE systems. Nevertheless, for small SIS breaking, the Berry curvature is highly localized at the valleys, and the local integral of the Berry curvature converges quickly to a non-zero quantized value \cite{ValleyContrasting, BilayerGraphene}. This local integral is often referred to as the valley Chern number $C_v$ of the $n^\textrm{th}$ band and it is defined as 
$2\pi C_v=\int \Omega _n (\mathbf{k}) d^2 \mathbf{k}$,
where the integral bounds extend to a local area around the valley. The right hand side of this equation is also referred to as topological charge. Previous studies in electronic systems \cite{ControlEdgeStates} have shown that this quantized value has important implications because it characterizes the bulk\textendash edge correspondence. In fact, the difference between the valley Chern numbers of the upper (or lower) bands of two adjacent lattices indicates the number of gapless edge states expected at the DW.

\begin{figure}[ht]
	\includegraphics[scale=0.6]{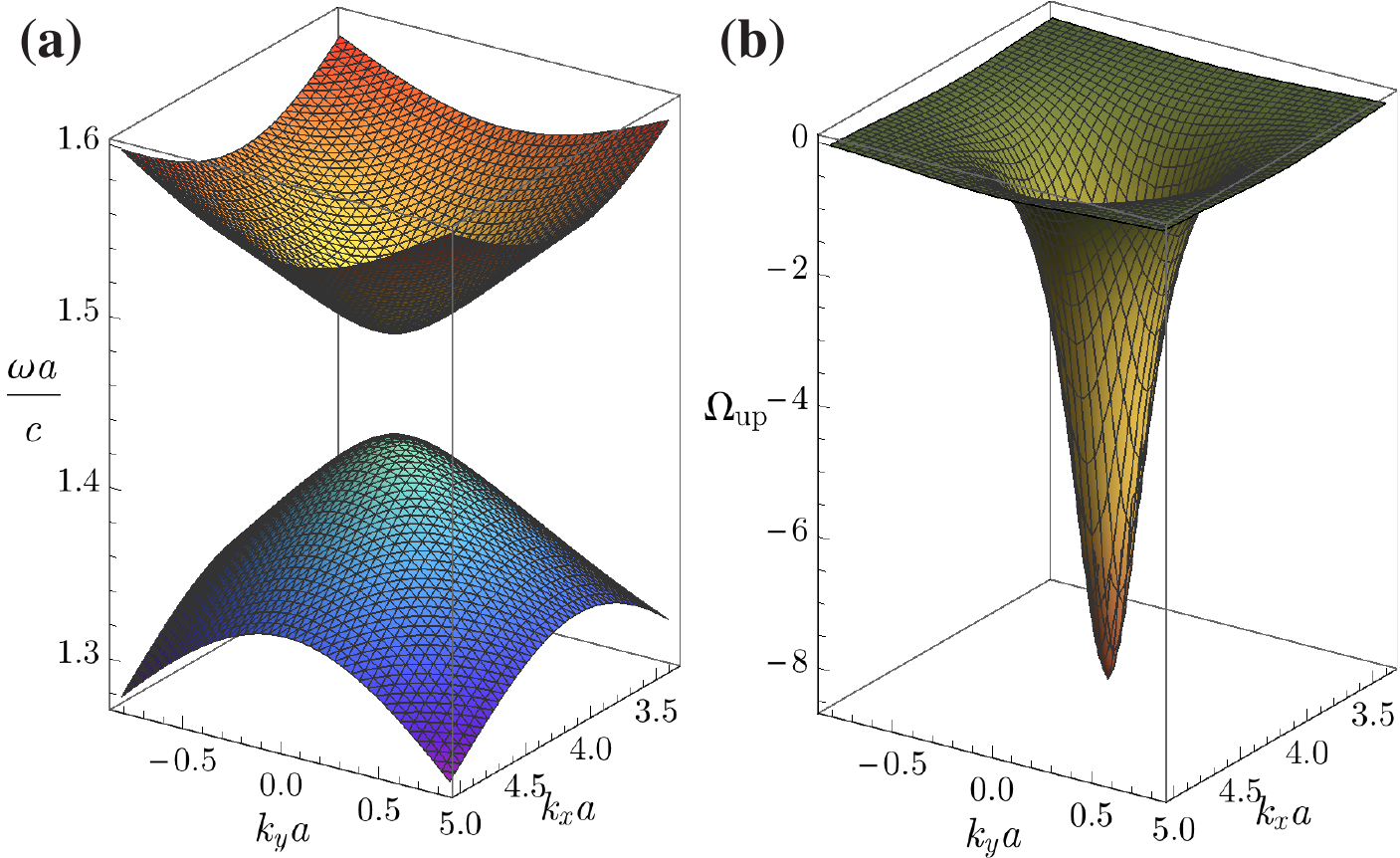}
	\caption{\label{fig:BC} (a) The dispersion of the flexural $A$ modes of the $\alpha$-lattice near the $\mathbf{K}$ valley. (b) The corresponding Berry curvature of the upper mode.}
\end{figure}

Numerical integration of the Berry curvature shows that each valley of either the $\alpha$- or $\beta$-lattices carries a topological charge of magnitude $\pi$. The two lattices clearly exhibit Berry curvatures of opposite signs. It follows that, for the upper mode of the valley $\mathbf{K}$, the valley Chern numbers are $- \frac{1}{2}$ for the $\alpha$-lattice and $+ \frac{1}{2}$ for the $\beta$-lattice. The difference $\lvert C^{(\alpha)}_v-C^{(\beta)}_v \rvert=1$ indicates the existence of a single gapless edge state at the DW between the $\alpha$- and $\beta$-lattices.
Fig.~\ref{fig:BC}(a) shows the calculated dispersion surfaces of the flexural modes of the $\alpha$-lattice near the valley $\mathbf{K}$.
Fig.~\ref{fig:BC}(b) shows the Berry curvature corresponding to the upper mode. Equivalently, the lower band carries a Berry curvature of opposite sign.
\section{Domain wall edge state}
\begin{figure}[ht]
	\includegraphics[scale=0.92]{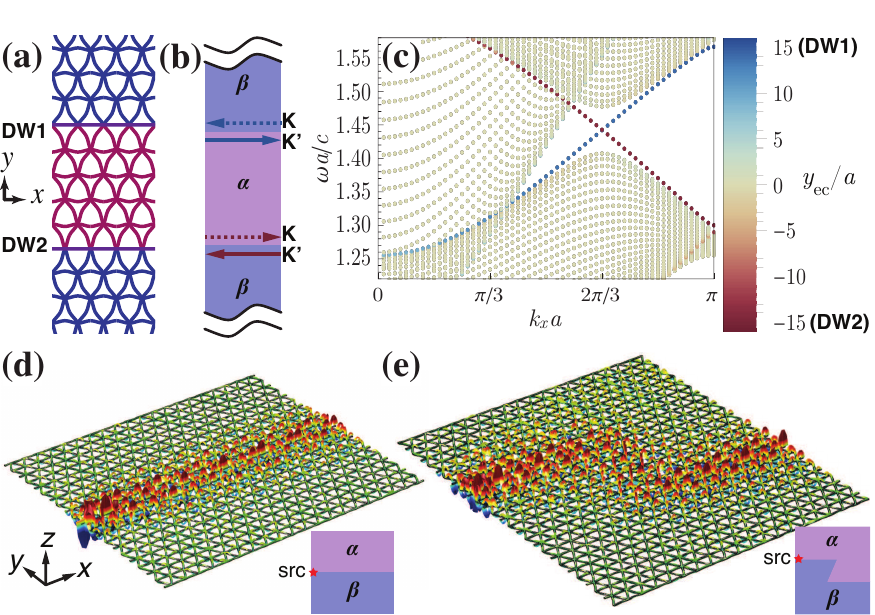}
	\caption{\label{fig:interface} (a) Lattice structure and (b) schematic illustration of the two possible DW configurations based on the $\alpha$- and $\beta$-lattices. Each DW allows propagation of an edge state for each valley index. (c) Dispersion curves in a superlattice ribbon containing 37 triangular subcells of $\alpha$-lattices stacked along the $y$-direction, with symmetric boundary conditions. The colorbar denotes the $y$-position of the centroid of the superlattice weighted by the strain energy density: modes in blue color are edge states at DW1, while modes in red indicates edge states at DW2. (d, e) Full field simulations of the edge states along a straight and an arbitrary-shaped DW.
	}
\end{figure}

There are two possible configurations of the DWs between the $\alpha$- and $\beta$-lattices, that is $\alpha$ above $\beta$ ($+y$) or viceversa (see Fig.~\ref{fig:interface} (a) and (b)). These two configurations are not equivalent, and are dominated by a mirror symmetry with respect to either DW1 or DW2. As a direct consequence of the mirror symmetry, the edge states propagating along the DWs can be either symmetric or anti-symmetric (with respect to the DW interface).

This symmetry was exploited in order to reduce the size of the finite element model. In particular, we built an $\alpha$-state superlattice ribbon made of 37 triangular subcells stacked along the $y$-direction (see the red part in Fig.~\ref{fig:interface} (a)) and having a length of about $64a$. Then, either symmetric or anti-symmetric boundary conditions along DW1 or DW2 were applied to simulate the presence of the $\beta$-state ribbon. The model was solved to get the Bloch eigenvalues and eigenvectors. Results show that the edge state at DW1 is anti-symmetric, while at DW2 it is symmetric. Fig.~\ref{fig:interface} (c) shows the dispersion relation of such a ribbon structure.

The colorbar denotes the $y$-position of the centroid of the superlattice weighted by the strain energy density. The blue color indicates edge states at DW1 while the red color indicates edge states at DW2. The two modes have equal and opposite group velocities but since they are supported by the two different edges of the lattice they cannot couple. In addition, if inter-valley mixing is neglected (due to large separation in $\mathbf{k}$-space between the $\mathbf{K}$ and $\mathbf{K}'$ points), these edge states are immune to back-scattering. Note that this assumption is well verified in the absence of short range disorder. We highlight that this dynamic behavior is effectively equivalent to
the quantum spin Hall effect in topological insulators \cite{ReviewKaneTI} if the spin index is replaced by the valley index.

In order to further characterize the egde modes, we performed full field numerical simulations. In particular, we simulated the steady-state response of two different DW shapes (i.e. straight and arbitrary) under a point excitation. Low-reflecting boundaries were used all around the model to suppress reflections. In both cases, results show that the edge states are well concentrated near the DWs and are guided along the wall itself (Fig.~\ref{fig:interface} (d),(e)).

\section{Semi-analytical approach}
Although the band structure and the valley Chern number were accurately calculated via a numerical approach, it is still useful to introduce a semi-analytical model to further analyze the effects of SIS-breaking on the phononic waveguide.
Starting from the (undeformed) reference lattice configuration which still preserves SIS, the $\mathbf{k}\cdot\mathbf{p}$ perturbation approach \cite{KPElastic} can be used to show that the linear dispersions at the valleys can be mapped to a massless Dirac equation. The method allows obtaining approximate solutions to the governing equations by using the degenerate modes as fundamental basis of an expansion process. In this method, the origin of the $\mathbf{k}$-axes is shifted to the degeneracy by letting $\mathbf{q} \equiv \mathbf{k}-\mathbf{K}$ (or $\mathbf{k}-\mathbf{K}'$) and $\nu \equiv \omega-\omega_0$. Hence, the problem of determining the dispersion relations $\nu$\textendash$\mathbf{q}$ is cast in the form of a $2 \times 2$ eigenvalue problem associated to the following Hamiltonian:
\begin{equation}
\mathcal{H}(\mathbf{q})= v_g \mathbf{q}\cdot \vec{\sigma}
\end{equation}
where $v_g$ is the group velocity, and $\vec{\sigma} \equiv (\sigma _1 , \sigma _2 , \sigma _3 )$ are Pauli matrices. The Hamiltonian maps to the massless Dirac equation associated with locally linear dispersion. By extracting the eigenstates from the previous simulations, we can numerically confirm that the group velocity $v_g=\pm0.138c$ obtained from the $\mathbf{k}\cdot\mathbf{p}$ method matches well with the tangent slopes of the exact dispersion data obtained from numerical calculations (inset Fig.~\ref{fig:bndstr} (a)).

When SIS is broken by the application of a pressure perturbation $\delta p$, the degeneracy at $\mathbf{K}$ (or $\mathbf{K}'$) is lifted, the modes become non-degenerate, and a gap opens up. It is well-known that breaking SIS introduces a $\sigma_3$-component into the Hamiltonian that can be expressed as \cite{ReviewKaneTI,Raghu}:
\begin{equation} \label{eq:massive}
\mathcal{H}(\mathbf{q})= v_g \mathbf{q}\cdot \vec{\sigma} +m \sigma _3
\end{equation}
It represents the Hamiltonian of a massive Dirac equation which we show applies also to our specific lattice structure:

Since the SIS-breaking perturbation is small, it can be expressed as a perturbation of the massless Hamiltonian,
\begin{equation} \label{eq:perturb}
\mathcal{H}(\mathbf{q})= v_g \mathbf{q}\cdot \vec{\sigma} +\vec{m} \cdot \vec{\sigma},
\end{equation}
where we assumed a general perturbation that potentially contains all the four Pauli matrix components ($\vec{\sigma}=[{\sigma_1,\sigma_2,\sigma_3}]$) as well as the identity matrix $\sigma_0$. At the \textbf{K} (or $\mathbf{K}'$) point ($\mathbf{q}=\mathbf{0}$), the Hamiltonian reduces to
\begin{equation} \label{eq:perturb_q0}
\mathcal{H}(\mathbf{0})= \vec{m} \cdot \vec{\sigma}.
\end{equation}
Let $\mathbf{u}^d_{1,2}$ be the two degenerate modes of the undeformed lattice and $\mathbf{u}^n_{1,2}$ be the two non-degenerate modes of the deformed lattice, at $\mathbf{q}=\mathbf{0}$. These modes are available from the numerical calculation of the band structure. By performing the inner product
\begin{equation} \label{innerproduct}
I_{ij}\equiv\underset{\textrm{unit cell}}{\int}\mathbf{u}^d_i\cdot \mathbf{u}^n_j d^3 \mathbf{r},
\end{equation}
it is seen that matrix $I$ is diagonal (or anti-diagonal, depending on the numbering of the modes), which implies that $\mathcal{H}(\mathbf{0}) = \vec{m} \cdot \vec{\sigma}$ is also diagonal. This means that the perturbation term can only contain $\sigma_0$ and $\sigma_3$ components, at most. In particular, the splitting of the eigenfrequencies is connected to the $\sigma_3$ component while the average upward shift of the bands (as a function of $\delta p$) is connected to the $\sigma _0$ component. Note that the $\sigma _0$ term does not affect either the eigenvectors or the topological properties, hence it can be omitted in the Hamiltonian. The above arguments confirm that the Bloch eigenvalue problem of the SIS-broken lattice can be mapped to the massive Dirac equation (Eq.~(\ref{eq:massive})).

The bulk dispersion is then given by $\nu(\mathbf{q})=\pm\sqrt{\lvert v_g\mathbf q \rvert^2+m^2}$ with a gap equal to $2\lvert m\rvert$. From the numerical results in Fig.~\ref{fig:fvspr} (b) we find that $\lvert m\rvert=0.000795 \textrm{MPa}^{-1}\delta p$, where $\delta p$ is expressed in MPa.

The corresponding Berry curvature, associated with the evolution of the eigenvectors in $\mathbf{k}$-space, has a distribution sharply peaked at the two Dirac points,
\begin{equation}\label{omega_b}
\Omega(\mathbf{q}) =\frac{1}{2}\tau m v_g^2 (\lvert \mathbf q\rvert^2 v_g^2+m^2)^{-\frac{3}{2}},
\end{equation} 
where $\tau=\pm$ allows simple labeling of the two valleys $\mathbf{K}$ and $\mathbf{K}'$. Integrating Eq.~(\ref{omega_b}) gives the valley Chern number $C_v=\frac{1}{2}\tau \textrm{sgn}(m)$. Under TRS, $m_\mathbf{K}=m_{\mathbf{K}'}$ for either a lattice $\alpha$ or $\beta$, and $m^{(\alpha)}=m^{(\beta)}$ at the same valley. Thus $\lvert C_v^{(\alpha)}-C_v^{(\beta)}\rvert =1$ is always true. Hence, we can conclude that there is one gapless edge state living on the DW connecting the two lattice $\alpha$ and $\beta$.

\section{Edge states at external boundaries}
\begin{figure}[ht]
\includegraphics[scale=0.8]{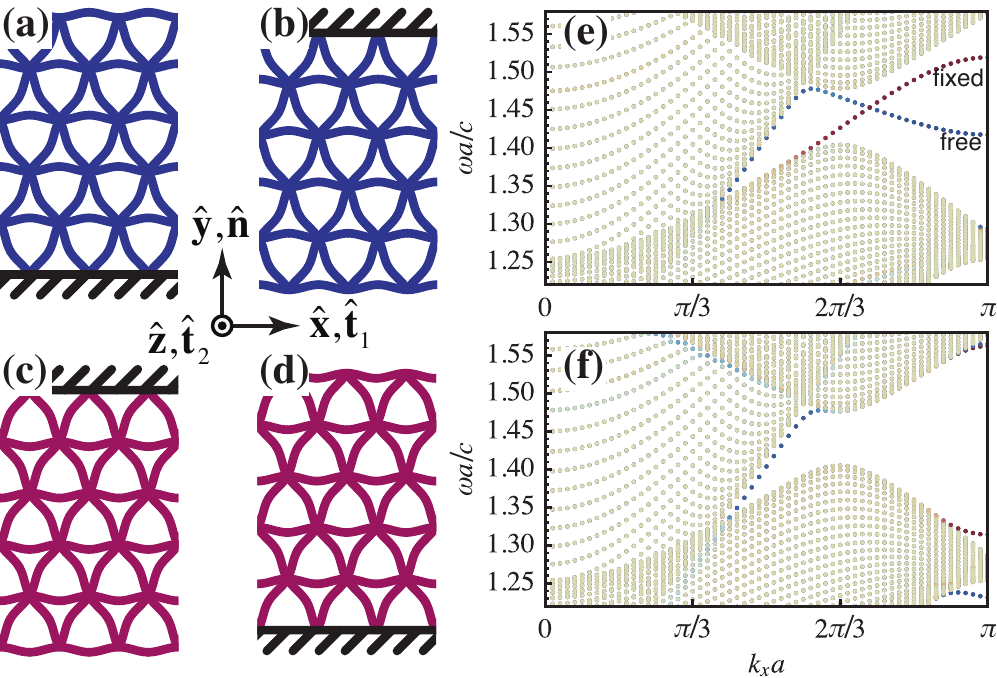}
\caption{\label{fig:fixedfree} (a-d) Schematic of the four different boundary configurations involving fixed and traction free conditions on the top $(+y)$ or bottom $(-y)$ boundaries of the $\alpha$- or $\beta$-ribbons. The configurations (a) and (c) yield two edge states associated with the dispersion shown in (e). The two boundary conditions in (b) and (d) do not yield topological edge states, as shown in (f).}
\end{figure}

The topological phase transition illustrated above and taking place between two lattice with broken SIS can be obtained also at the boundary of an individual lattice when proper boundary conditions are applied. The two specific boundary conditions considered in this study are traction-free and fixed. Each boundary condition can be applied on the top $(+y)$ and bottom $(-y)$ boundaries of the ribbon (either in $\alpha$- or $\beta$-state), hence resulting in a total of eight possible configurations (Fig.~\ref{fig:fixedfree} (a)-(d)). However, since the $\alpha$- and $\beta$-lattices are inverted images of each other, a given boundary condition on the edge of an $\alpha$-ribbon corresponds to the inverted image of the same boundary condition on the opposite edge of a $\beta$-ribbon, therefore yielding only four different edge dispersions
(for example, see the pairs (a)/(c) and (b)/(d)) in Fig.~\ref{fig:fixedfree}.
Note that, although they have the same edge dispersion, the valley indexes interchange. 

Not all the four groups of boundary conditions yield edge modes. Only the two edges of the ribbon $\beta$ in Fig.~\ref{fig:fixedfree} (a) (or the edges of ribbon $\alpha$ in Fig.~\ref{fig:fixedfree} (c)) have propagating edge states, as shown in Fig.~\ref{fig:fixedfree} (e). The other two groups (Fig.~\ref{fig:fixedfree} (b) and (d)) do not support edge states (Fig.~\ref{fig:fixedfree} (f)). Note that in both cases (Fig.~\ref{fig:fixedfree} (e) and (f)) there are localized modes near the free edge whose dispersion curves run along the bulk bands corresponding to the faster flexural branch. These modes have no topological significance and they are effectively surface-like waves with depth of the same order of the wavelength, which will be discussed in detail in a later section. These results can be explained by identifying the similarity between the dynamic behavior imposed by the boundary conditions and the DWs, respectively.

Recall that DW1 supports an anti-symmetric edge state. Anti-symmetry requires $\mathbf{u}\cdot\mathbf{t}=0$ on the interface, where $\mathbf{t}$ is any tangent vector to the interface, therefore having the direction of either $\hat{\mathbf x}$ or $\hat{\mathbf z}$. To satisfy the antisymmetry condition $u_x$ and $u_z$ must be zero, while the $u_y$ component are non-zero. The $A$ modes, under consideration in this study, exhibit particle displacement mostly dominated by the $u_z$ component. When a fixed boundary condition is imposed along the edge, the particle displacement \textbf{u} is set to zero therefore resembling the behavior of the anti-symmetric interface. On the other hand, DW2 supports symmetric edge states. At the interface we have $\mathbf{u}\cdot\hat{\mathbf{y}}=0$, without restriction on $u_z$. Similarly to the discussion above, the traction free boundary condition will allow $u_z$ therefore behaving more closely to a symmetric interface.

As a result, each of the fixed boundaries in Fig.~\ref{fig:fixedfree} (a) and (c) (note that this figure is equivalent to Fig. 5 in the main text; here reported for convenience) is dynamically equivalent to half of the DW1. This boundary configuration supports edge states similar to those observed along DW1, but without connecting the two continuous bulk bands (i.e. the red modes in Fig.~\ref{fig:fixedfree} (e)). Similarly, each of the free boundaries in Fig.~\ref{fig:fixedfree} (a) and (c) supports an edge state that does not cross the bandgap. On the contrary, the other two groups of boundary conditions (Fig.~\ref{fig:fixedfree} (b) and (d)) break the similarity with the DWs, therefore the edge states are strongly suppressed.

\begin{figure}[ht]
\includegraphics[scale=1]{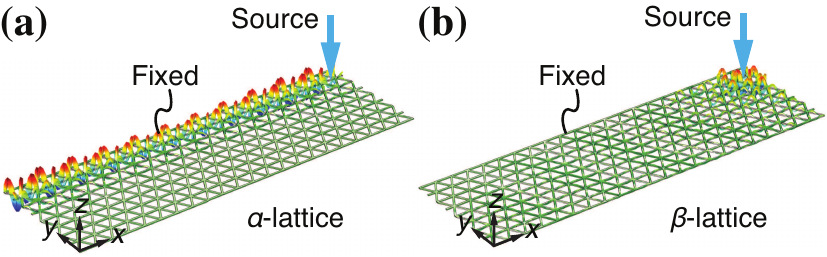}
\caption{\label{fig:freq} Full field numerical simulations showing that (a) edge states can be excited on the fixed edge on the top of an $\alpha$-ribbon, while (b) they cannot be excited at the same position in a $\beta$-ribbon.}
\end{figure}

Full field numerical simulations confirmed this behavior and showed that edge states can be excited on the fixed edge on the top of an $\alpha$-ribbon, but they cannot be excited under equivalent conditions in a $\beta$-ribbon (see Fig.~\ref{fig:freq} (a) and (b)). These two conditions correspond to the fixed edges in Fig.~\ref{fig:fixedfree} (c) and (b), respectively.

\section{lossless edge states}

The results above deserve some further discussion to clarify why the edge states can be excited independently from the bulk states existing at the same frequency. Due to the lack of a full bandgap, when a point excitation is placed at the domain wall, both bulk and edge states can potentially be excited. However, two factors contribute to make the edge states predominant compared to the bulk states. First, the bulk states amplitude decays as $1/r$ from the source, while the edge state amplitude remains constant. Second, since the lattice is inhomogeneous, the mode shapes of either the bulk or the edge states have non-uniform distribution over the unit cell. If a point source is located at the point where the edge state has the largest displacement amplitude, most of the input energy will be injected into the edge rather than the bulk state.
In Fig.~\ref{fig:freq} (b) if the edge state was suppressed, the long wavelength ripples of the bulk state would be visible.
In summary, both states exists but their large separation in $k-$space allows selective excitation of the edge states. 
In a similar way, if a distributed source was used other than a point source, the valley injection could still be achieved by targeting a selected wavenumber that supports strong edge modes. This results could be achieved by using excitation methods such as comb transducers.

\begin{figure}[ht]
\includegraphics[scale=1]{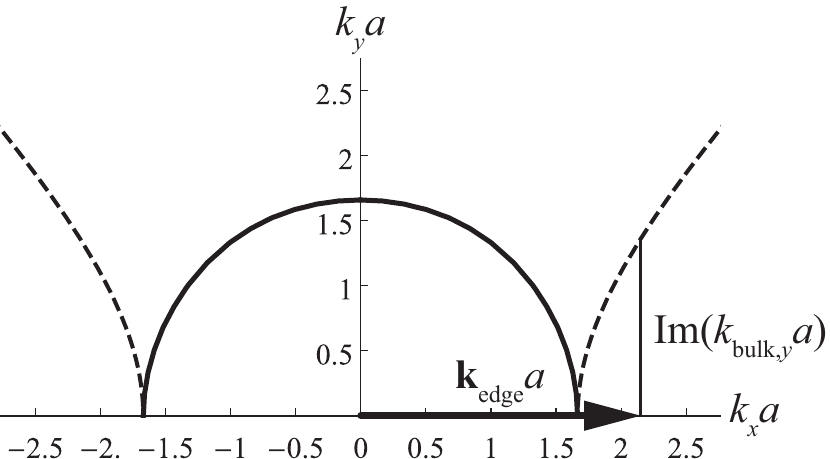}
\caption{\label{fig:slowness} The equi-frequency contour of the bulk state having the $x$-axis aligned with the edge (DW or boundary) at the edge state's frequency. By imposing the conservation of the wavevector component parallel ($k_x$) to the interface and providing that the bulk state is faster that the edge state, results in scattered waves towards the bulk that are evanescent in the normal ($k_y$) direction and therefore the edges state is non-leaky.}
\end{figure}

In addition, it can be shown that the edge state propagates without loss. At the frequency of the edge state, we calculate the equi-frequency contour (with the $x$-axis is aligned with either the edge, the DW, or the boundary) of the faster bulk state which is indicated by the solid curve in Fig.~\ref{fig:slowness}. The curve shows an almost circular profile for the propagating bulk state while the dashed line indicates the imaginary $k_{\mathrm{bulk},y}a$ component corresponding to an arbitrary large $k_xa$ beyond the circle. This is equivalent to a conventional slowness diagram, but different by a constant factor $a/\omega$. When the edge state propagates and induces bulk wave scattering, the $k_xa$ component must be conserved, that is $\lVert \mathbf{k}_{\mathrm{edge}} \rVert = k_{\mathrm{bulk},x}$. Clearly since the bulk state is faster, there exists no $\mathbf{k}_\mathrm{bulk}$ on the solid curve satisfying this condition. The scattered bulk state then has a complex wavevector with purely imaginary $k_y$ component as illustrated graphically in Fig.~\ref{fig:slowness}. This means that the scattered bulk wave is evanescent along $y$-direction and travels with the same speed as the edge state along $x$-direction. As a result, the edge state is lossless and it does not necessarily require a bulk band gap as long as the bulk state is faster (in terms of phase velocity) than the edge state at the same frequency.

\section{Tuning the boundary edge states}
In addition to the effect of the boundary conditions, adjusting the pressure of the outermost cells (those close to the boundary) can significantly tune the edge state dispersion. Yao et al. \cite{ControlEdgeStates} showed that on a graphene system with staggered sublattice potential, the edge states dispersion can be controlled by the on-site energy on the boundary cells. Similarly, in our system the phononic edge states dispersion can be manipulated by tuning the pressure of the outermost row of lattice. From the dispersion curve (Fig.~\ref{fig:fixedfree} (e)), we observe that the edge state develops from the bulk band near the valley and as $k_x$ increases the edge state localizes more sharply on the edge. At $k_x a = \pi$, the energy is mostly confined in the outermost row of the lattice, therefore controlling the edge state requires tuning the outermost cell properties.
For our phononic waveguide, a general trend can be established observing that a positive pressure on the outer cells tends to increase the edge state frequency at $k_x a = \pi$, while a negative pressure tends to decrease it. This suggests that the edge state dispersion can be tailored by controlling the local pressure in the edge cells. 

\begin{figure}
\includegraphics[scale=0.8]{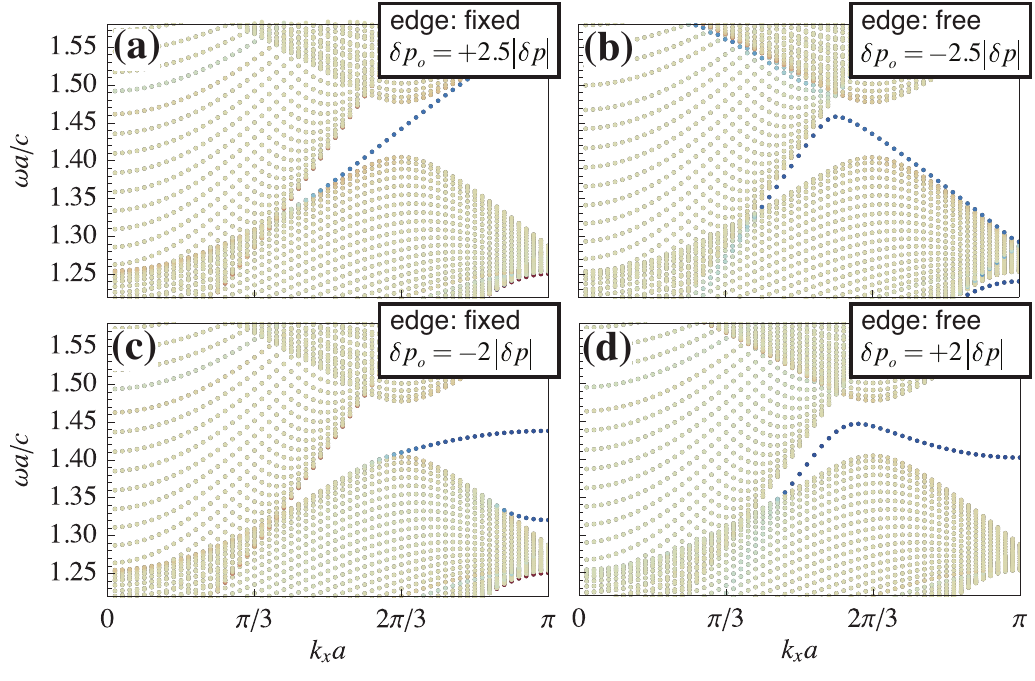}
\caption{\label{fig:fixedfree_tuned} (a, b) Edge states at the fixed and free edge when the outer cell pressure is set to $\delta p_o = +2.5 \lvert \delta p \rvert$ and $-2.5 \lvert \delta p \rvert$, respectively. The edge states can be tuned to bend towards either the top or the bottom bulk bands. (c, d) Edge states at the fixed and free edge when the outer cell pressure is set to $\delta p_o = -2 \lvert \delta p \rvert$ and $+2 \lvert \delta p \rvert$, respectively. The edge states become flat bands.}
\end{figure}

Numerical results in Fig.~\ref{fig:fixedfree_tuned} show that
a combination of boundary conditions and local pressure allows tuning the edge states to either a partially gapped band, a gapless band, or a flat mode. As an example, by setting the outer cell pressure to $\delta p_o = +2.5 \lvert \delta p \rvert$ near the fixed edge (cf. Fig.~\ref{fig:fixedfree} (c)) the dispersion curve of the edge state bends up towards the top bulk band (Fig.~\ref{fig:fixedfree_tuned} (a)).
Similarly, by setting the outer cell pressure to $\delta p_o = -2.5 \lvert \delta p \rvert$ near the free edge (cf. Fig.~\ref{fig:fixedfree} (a)) the dispersion curve bends down towards the bottom bulk band (Fig.~\ref{fig:fixedfree_tuned} (b)). Flat bands can also be obtained (Fig.~\ref{fig:fixedfree_tuned} (c) and (d)) by applying $\delta p_o = 2 \lvert \delta p \rvert$ and $-2 \lvert \delta p \rvert$ in the previous two cases, respectively. For the two cases supporting the gapless edge states crossing the gap, the pressure applied on the outer cells has different sign from the bulk lattice.
This situation is somewhat equivalent to introducing a DW on the boundary and it results again in the occurrence of gapless edge states due to the contrast between bulk topological charges.

\subsection{Discussion of the non-topological edge state}
\begin{figure}[ht]
\includegraphics[scale=1]{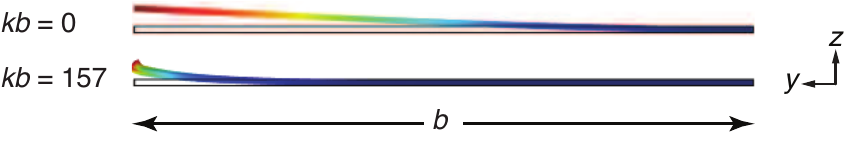}
\caption{\label{fig:fin_modeshape} Cross-section view of the mode shapes of a fin-like elastic waveguide with thickness $t$ and width $b$, at $kb=0$ and $kb=157$. The fundamental flexural mode develops into a surface wave as $kb\gg 1$.}
\end{figure}

\begin{figure}[ht]
\includegraphics[scale=1.5]{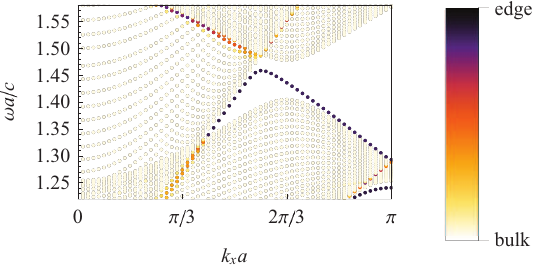}
\caption{\label{fig:coupling} Results of Fig.~\ref{fig:fixedfree_tuned} (b) plotted using a nonlinear color scheme based on the particle displacement. This plot clarifies the interaction between the two edge modes. The gapless topological edge state and the non-topological edge state meet at the upper left of the \textbf{K} point, and experience a strong repulsion which indicates mode hybridization. When both modes enter the continuous bulk band, they become leaky waves due to the coupling with the bulk modes.}
\end{figure}

These results deserve some additional discussion concerning the dispersion of the traction-free edge (Fig.~\ref{fig:fixedfree} (e), (f) and Fig.~\ref{fig:fixedfree_tuned} (b), (d)). Note that the dispersion curves highlight the existence of a localized mode near the free edge, which develops along the bulk band corresponding to the faster flexural branch.
These parts have no topological significance and they are effectively surface-like waves with depth of the same order of the wavelength.
To clarify this aspect further, consider a fin-like elastic waveguide with thickness $t$ and width $b$. The fin can be thought as a continuum version of our phononic ribbon without the lattice structure. It is well-known that as $kb\gg 1$ the fundamental flexural mode develops into a surface wave (Fig.~\ref{fig:fin_modeshape}). This is similar to the $S_0$ and $A_0$ Lamb modes evolving into Rayleigh surface modes as $kt\rightarrow \infty$. Therefore, this edge mode has no topological significance as confirmed also by the zero Berry curvature being associated with the bulk bands. Nevertheless, there is strong coupling between this mode and the topological edge states propagating along the same traction-free edge. To clarify the interaction between these two edge modes, we can plot again the results of Fig.~\ref{fig:fixedfree_tuned} (b) with an emphasized nonlinear color scheme in Fig.~\ref{fig:coupling}. The gapless topological edge state and the non-topological edge state meet at the upper left of the cone, and there is a strong level repulsion indicating hybridization. When they enter the continuous bulk band of the other, they become strongly leaky due to the coupling with bulk modes.

\section{Selective valley-injection and one-way propagating edge states}
\begin{figure}[ht]
	\includegraphics[scale=1.05]{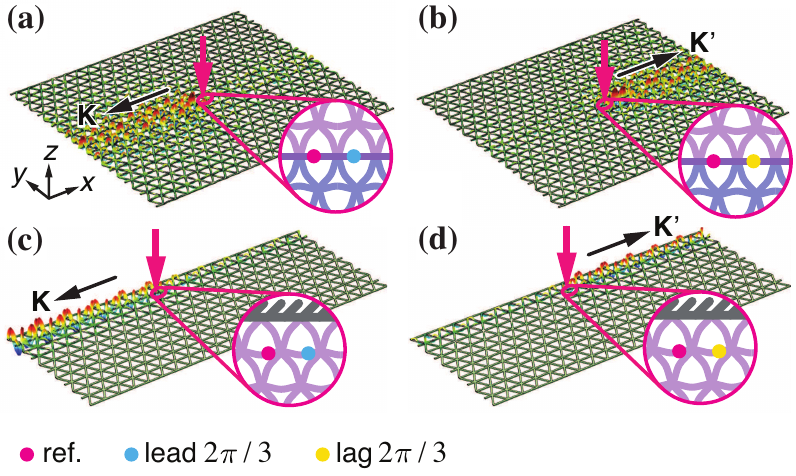}
	\caption{\label{fig:directional} Full field numerical simulations showing the ability to trigger a specific uni-directional edge state (either around the \textbf{K} or $\mathbf{K}'$ point) along either (a,b) a DW or (c,d) a fixed edge. The insets show that by selecting a proper two-point excitation with appropriate phase difference (matched to the targeted Block mode) accurate valley injection can be achieved.}
\end{figure}

Finally we show that, despite the TRS remains intact, the two valley-dependent edge states can be excited individually, therefore giving rise to one way propagation. Since the two valley edge modes are largely separated in momentum space, the coupling between them is vanishingly small. This means that \textit{valley-injection} can be achieved by using a source having stronger coupling with one of the two valleys. As an example, referring to the Bloch modes, two point sources distanced by $a$ and having a phase difference $\pm 2\pi /3$ can selectively excite a specific mode.
Full field numerical results are shown in Fig.~\ref{fig:directional} (a, b) for the DW and in Fig.~\ref{fig:directional} (c, d) for the fixed boundary. Again, low reflecting boundaries were used all around the model to suppress reflections.

\section{Conclusions}
In conclusion, we presented the design of a tunable topological elastic phononic waveguide based on the acoustic analog of the QVHE. The lattice structure exploited SIS-breaking while preserving TRS, hence resulting in a weak topological acoustic material. SIS-breaking was induced by tunable elastic deformation of the original lattice structure while the topological phase transition was achieved by contrasting ribbons having different topological charges. This configuration could produce clear edge states at the domain wall interface. In a similar way, the proposed design was shown to be able to achieve topological edge states even in a single lattice configuration when in presence of specific boundary conditions, and the edge states could be efficiently tailored by tuning the pressure in the outermost edge cells. We also showed that, despite TRS being preserved, selective valley injection could be effectively achieved by a two-point excitation therefore giving rise to uni-directional edge modes. We also note that, in presence of smooth disorder, the inter-valley mixing is fairly weak which allows for back-scattering immune edge states. 
\begin{acknowledgments}
The authors gratefully acknowledge the financial support of the Air Force Office of Scientific Research under the grant YIP FA9550-15-1-0133.
\end{acknowledgments}

\bibliography{Ref}
 
\end{document}